\documentstyle[psfig]{article}
\def\bc{\begin{center}}
\def\ec{\end{center}}
\newcommand{\beq}{\begin{equation}}
\newcommand{\eeq}{\end{equation}}
\newcommand{\beqa}{\begin{eqnarray}}
\newcommand{\eeqa}{\end{eqnarray}}
\begin{document}
\begin{center}
{\bf {\LARGE  AGAPE, a microlensing search in the direction of M31: 
    status report}}
\\
~\\
{\large R. Ansari$^a$, M. Auri{\`e}re$^b$, P. Baillon$^c$, A. Bouquet$^{d,f}$, 
G. Coupinot$^b$, \\C. Coutures$^e$, C. Ghesqui{\`e}re$^f$,
Y. Giraud-H{\'e}raud$^f$, P. Gondolo$^{d,g}$,J. Hecquet$^b$,\\ 
J. Kaplan$^{d,f}$, Y. Le Du$^f$, A.L. Melchior$^f$, M. Moniez$^a$,
J.P. Picat$^b$,\\ 
and G. Soucail$^b$ }
\vspace{1cm}

$^a$ {\em LAL, Universit{\'e} Paris Sud, Orsay, France}, \\
$^b$ {\em Observatoire Midi-Pyr{\'e}n{\'e}es Bagn{\`e}res de Bigorre et Toulouse, France},\\
$^c$ {\em CERN, Gen{\`e}ve, Switzerland}, 
$^d$ {\em LPTHE, Universit{\'e}s Paris 6 et 7, France}, \\
$^e$ {\em DAPNIA, CEN Saclay, France}, 
$^f$ {\em LPC Coll{\`e}ge de France, Paris, France.} \\
$^g$ {\em Theoretical Physics, University of Oxford, United Kingdom}
\vspace{.5cm}

Presented at the second Workshop : ``The dark side of the Universe :
experimental efforts and theoretical framework'' Roma 13\,-14 November 1995, by J. Kaplan.
\end{center}
\vspace{.3cm}
LPC 96-04/conf
\vspace{.3cm}

The M31 galaxy in Andromeda is the nearest large galaxy after the Small and Large Magellanic
Clouds. It is a giant galaxy, roughly 2 times as large as our Milky Way, and has its own halo.
As pointed by some
of us {\small\cite{BBGK,BBGK1}} and independantly by
A. Crotts {\small\cite{crotts}},  M31 provides a rich field of stars  to search for MACHO's
in galactic halos by gravitational microlensing {\small\cite{pacz}}. M31 is a
target complementary to the Magellanic Clouds used by the current
experiments {\small\cite{MACHO,EROS}}. 
It is complementary in that it allows to probe the halo of our galaxy
in a direction very different from that of the LMC. Moreover, the fact that M31 has 
its own halo and is tilted with respect to the line of sight provides a
very interesting signature {\small\cite{crotts}}~: assuming an approximately spherical halo for M31, the  far side
of the disk lies behind a larger amount of M31 dark matter, therefore more microlensing
events are expected on the far side of the disk. Such an asymmetry could not be faked by
variable stars.

In other words, M31 seems very appropriate to detect brown dwarfs through microlensing.
However, as very few stars of M31 are resolved, we had to develop a new approach to look for
microlensing by monitoring the pixels of a CCD, rather than individual
stars {\small\cite{BBGK,BBGK1}}. The AGAPE collaboration has set out to implement
this idea.\\

{\bf \large Monitoring pixels}

In the case of a crowded field such as M31, the light flux $F_{\mbox{\rm pixel}}$ on a pixel comes from  the 
many stars in and around it, plus the sky background. The light
flux of an  individual
star, $F_{\mbox{\rm star}}$,  is spread among all pixels of the seeing spot and only a
fraction of this light, $F_{\mbox{\rm pixel}} = \left\{  \mbox{\rm seeing fraction}
\right\} \times \ F_{\mbox{\rm star}}$, reaches the central pixel.  If the star luminosity is amplified by a factor $A$, the pixel flux
increases by :
\beq
\Delta F_{\mbox{\rm pixel}} = (A-1) \ \left\{ \mbox{\rm seeing fraction}
 \right\} \ F_{\mbox{\rm star}}\,.
\eeq
The amplification of the star
luminosity allows an event to be detected if 
the flux on the brightest pixel rises sufficiently high above its rms
fluctuation $\sigma_{\mbox{\rm pixel}}$ :
\beq
\Delta F_{\mbox{\rm pixel}} > Q\ \sigma_{\mbox{\rm pixel}}\,. \label{detect}
\eeq
Typically, in our simulations, we require $Q$ to be larger than 3 during 3 consecutive
exposures and larger than 5 for at least one of them.

All our simulations {\small\cite{BBGK,BBGK1,munich}} indicate that 
the statistics should be significant if the relative fluctuation of
the current pixels (i.e. where no particular activity occurs) can be
made sufficiently small. This crucial point is the first step of our
data analysis and is discussed in detail below.\\

{\bf \large Status of the analysis}

\begin{figure}[t!]
\begin{minipage}[t]{.5\textwidth}
  {\psfig{figure=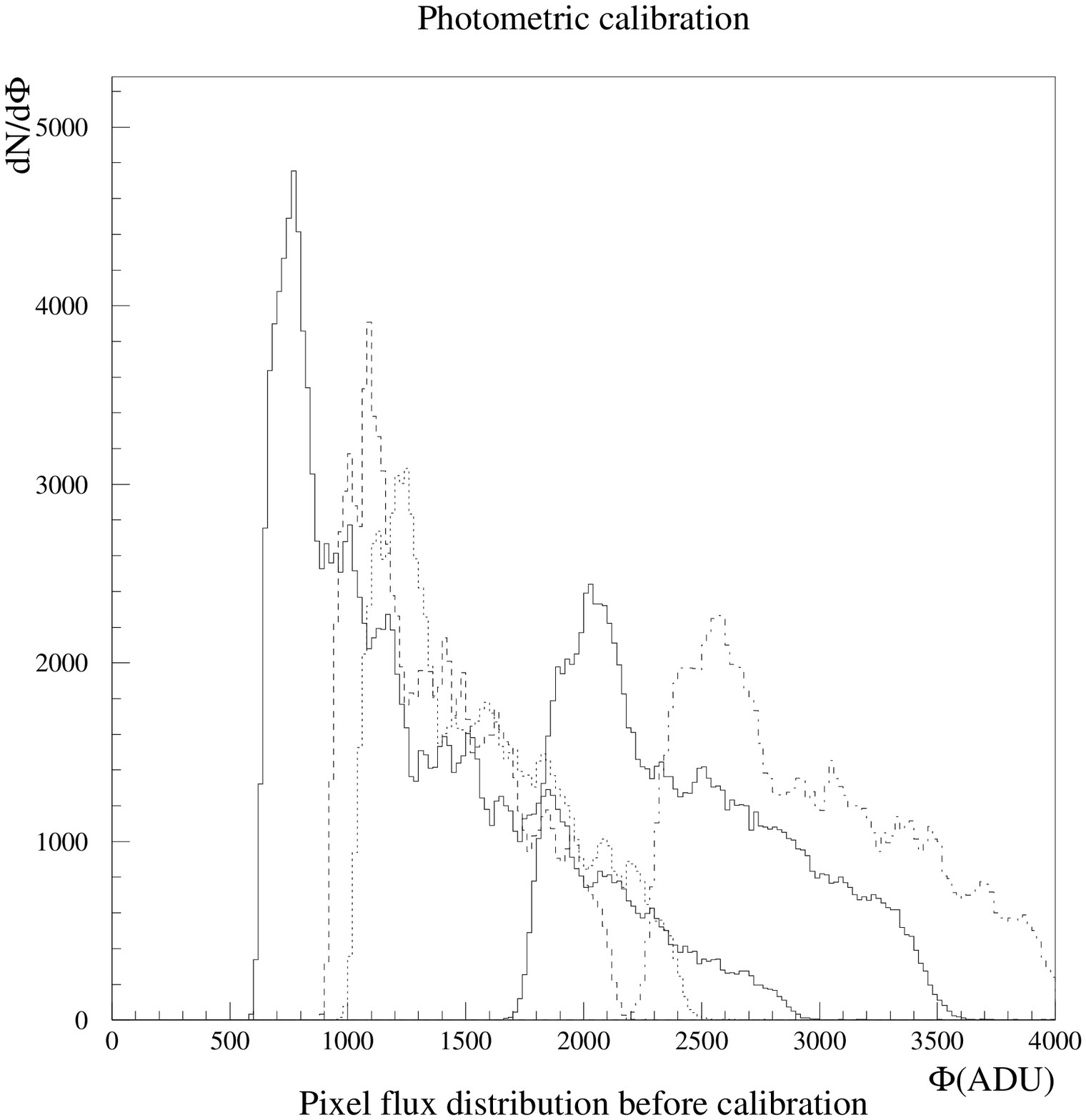,width=\textwidth}}
  \begin{center} a \end{center}
\end{minipage}\hfill
\begin{minipage}[t]{.5\textwidth}
  {\psfig{figure=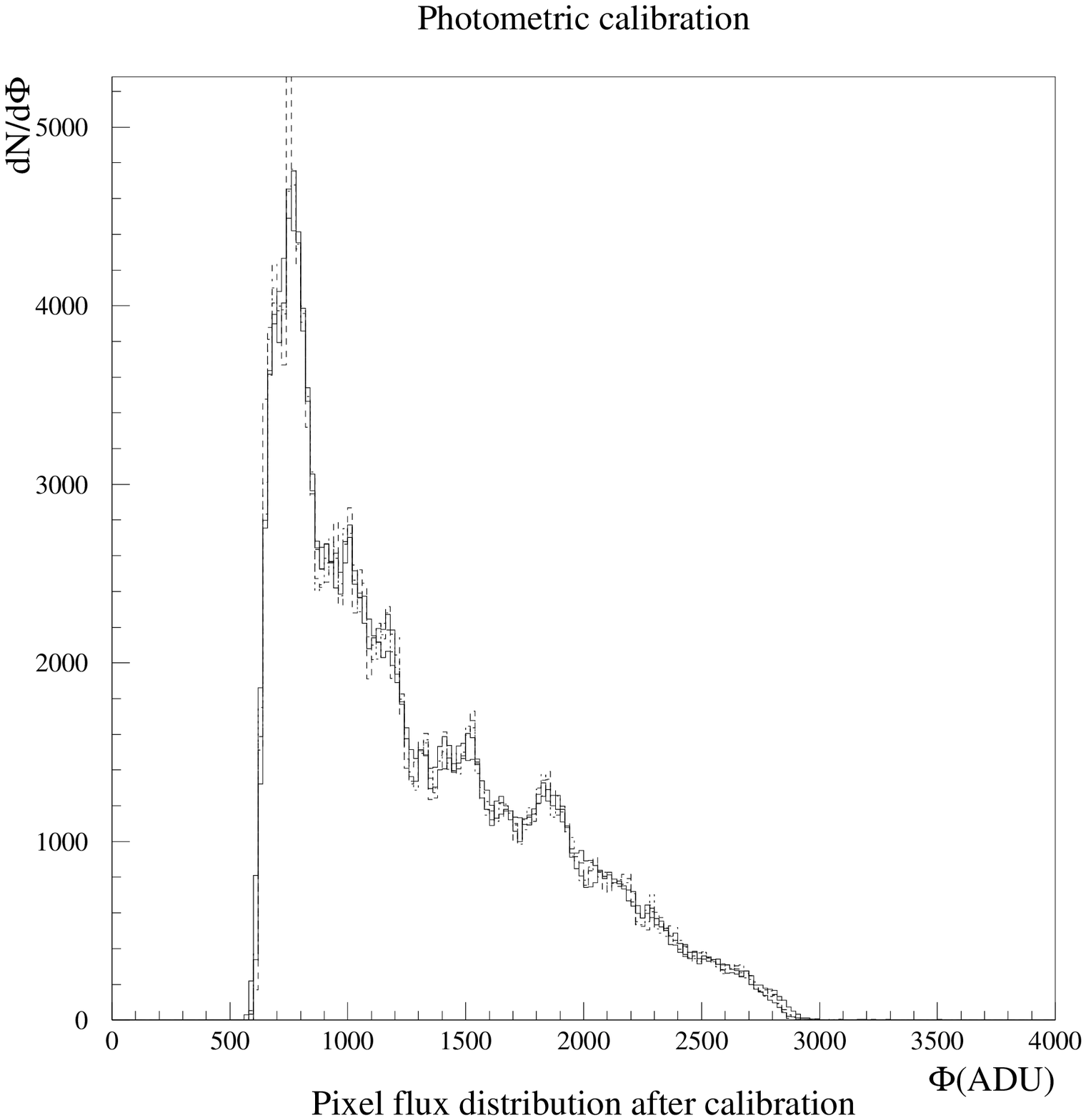,width=\textwidth}}
  \begin{center} b \end{center}
\end{minipage}\hfill
\caption{The matching of pixel histograms before (a) and after (b)
  photometric alignment
  \label{xxx}}
\end{figure}
We took data during 60 nights of observation on the 2 meter ``Bernard
Lyot'' telescope at Observatoire du Pic du Midi in the French
Pyr{\'e}n{\'e}es, from September 29 to November 24 in 1994, and 93 
nights, from July 28 to Dec 31 in 1995. Of these 153 nights, only 61
came out with good weather.
The field regularly covered was $8' \times 8'$,with 4
exposures on a $800 \times 800$ part of a thin Tektronix CCD camera
with pixels $0.3\,''$ wide. Two other $4' \times 4'$ fields were occasionally
covered.

Images have been taken with both red (Gunn) and blue (Johnson)
filters, but less regularly in blue. We have not yet started the
analysis of the blue frames. \\ 

\textbf{Geometrical alignment}.
A definite pixel never points exactly in the same direction of the sky on two
different exposures. Successive images have therefore to be realigned 
geometrically by software, in order that the light curve of a pixel really
represents the light curve of a definite region of the sky. To this
aim, we match the positions of bright stars between the current image
and a reference image, using an adaptation to our case of the program PEIDA
devised by the EROS collaboration {\small\cite{PEIDA}}. The precision
of the geometrical realignment thus obtained is better than 0.3 pixel
(0.1$\,''$). Such a precision is fully satisfactory, as we construct
our light curves on super-pixels of size $5\, \mathrm{pixel} \times 5
\,\mathrm{pixel}$ or larger.\\  

\textbf{Photometric alignment}.
\begin{figure}[ht!]
  \bc \begin{minipage}[t]{\textwidth}
    {\psfig{figure=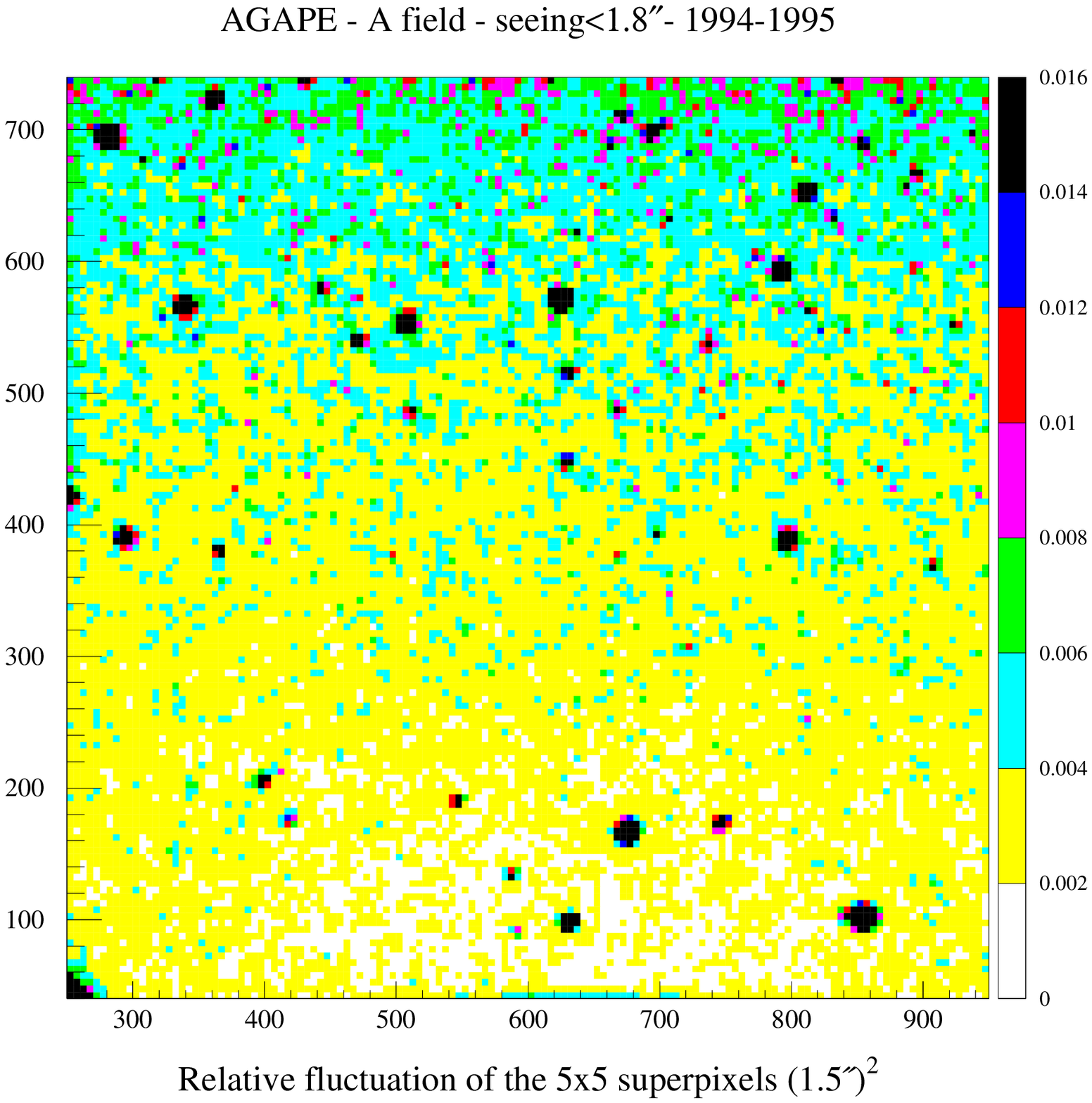,width=0.75\textwidth}}
  \end{minipage}\ec
  \vspace{-0.5cm}
  \caption{The relative fluctuation on field A \label{YYY}}
\end{figure}
The sky background light and the atmospheric
absorption differ from one picture to the other. Before any
photometric follow-up, it is necessary to correct for these
differences. We do that by matching the dispersion and the mean value
of the histogram of pixel intensities between each frame and a
reference one. This method works well in this
case because the local luminosity gradient in M31 largely supersedes
all other sources of dispersion. 
Figure \ref{xxx} shows how well 
pixel histograms, that look very different before treatment, coincide
up to small structures after a renormalization by only two
parameters~: an overall multiplicative factor for the atmospheric 
absorption and an additive constant for the sky background.
To check the quality of the photometric alignment obtained, we
evaluated the relative intensity fluctuation of each super-pixel
among all exposures.
Figure~\ref{YYY} shows a map of this relative dispersion for $5 \times
5$ super-pixels for one of our fields. We see that this dispersion
does not exceed 0.5\% on most of the field, which is around twice
the photon noise. 
\begin{figure}[h!]
  \begin{minipage}[t]{.5\textwidth}
    {\psfig{figure=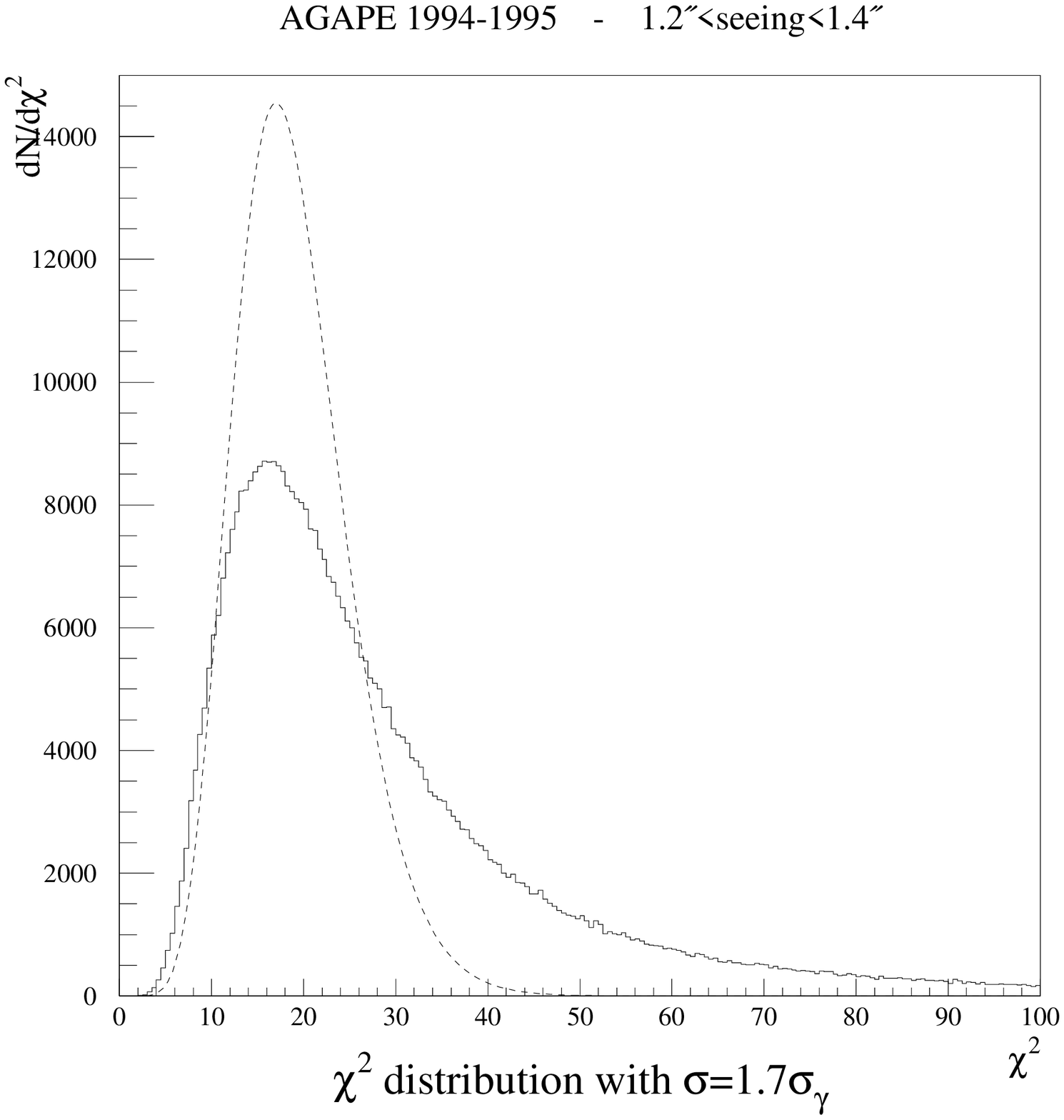,width=\textwidth}}
  \end{minipage}\hfill
  \begin{minipage}[t]{.5\textwidth}
    {\psfig{figure=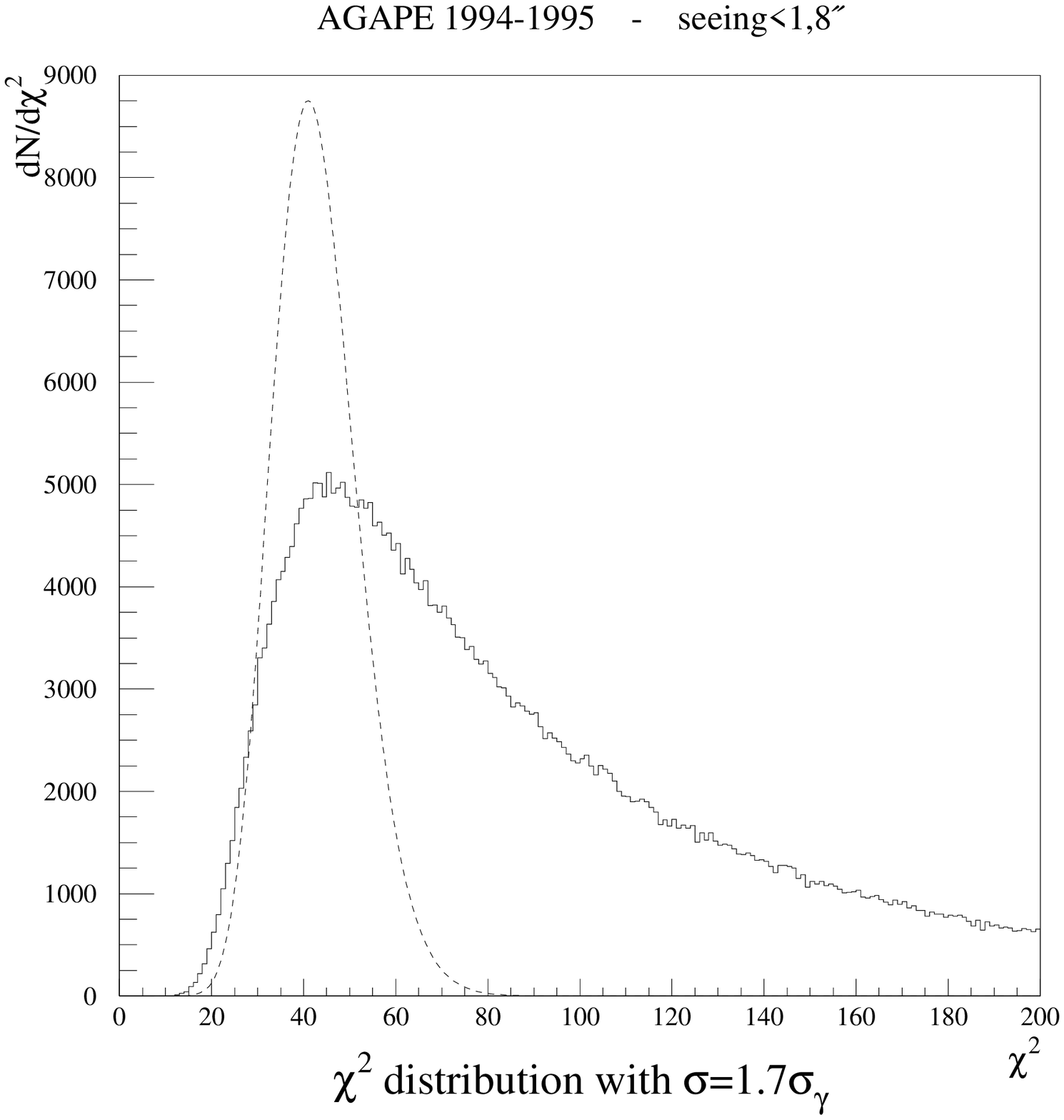,width=\textwidth}}
  \end{minipage}\hfill
  \caption{The relative fluctuation on field A 
    \label{ZZZ}}
\end{figure}
This can be looked at in a different way (figure~\ref{ZZZ}): after constructing a light
curve for each super-pixel, one can compute along each light curve the
$\chi^2$ of the intensity compared with its average. 
On figure \ref{ZZZ}, we display for two different seeing 
intervals the distribution of this $\chi^2$. The error $\sigma$
entering the $\chi^2$ is chosen in such a way that the maximum of the
distribution coincides with that of the ideal poissonnian $\chi^2$
distribution. The true distibutions show non poissonnian
tails. Clearly there are non gaussian contributions to the fluctuations
and a comparison between figure
\ref{ZZZ}a and~\ref{ZZZ}b shows that they are largely due to seeing
variations. Further work is in progress to cope with seeing
variations.  \\  

\textbf{Present results}.

Let us examplify the kind of results we get by the light curve of a 
\begin{figure}[h!]
  \bc  \begin{minipage}[t]{\textwidth}
    {\psfig{figure=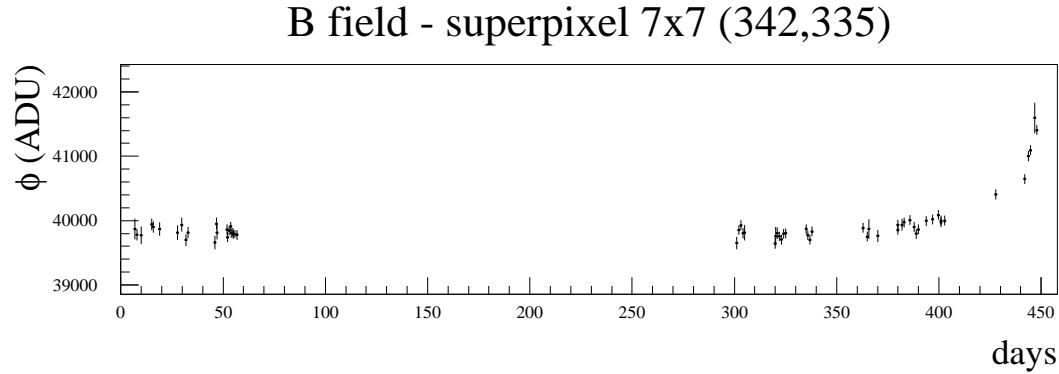,width=\textwidth}}
  \end{minipage} \ec
\vspace{-0.5cm}
  \caption{A typical light curve}\label{WWW}
\end{figure}
\begin{figure}[h!]
  \bc  \begin{minipage}[t]{\textwidth}
    {\psfig{figure=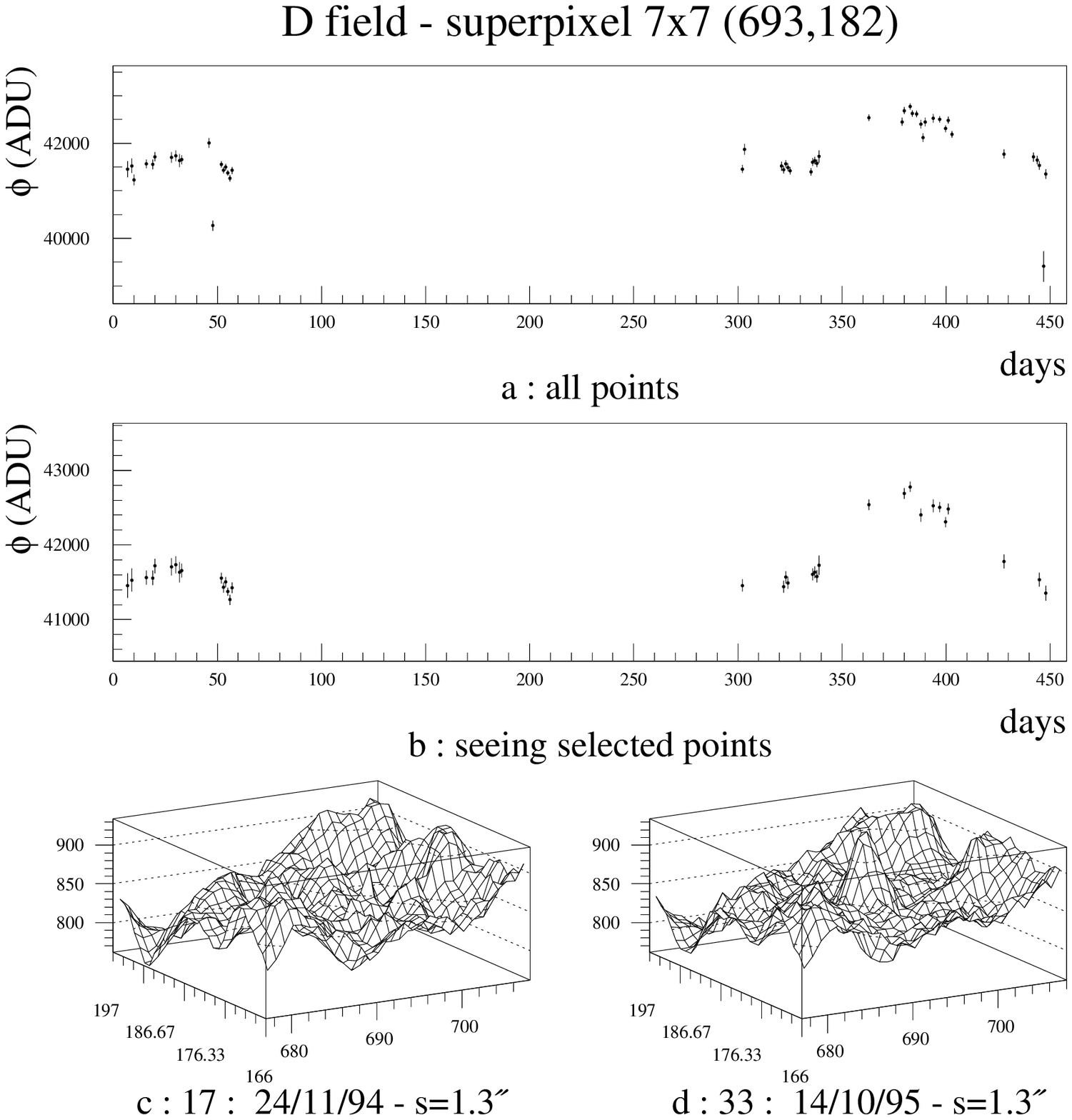,width=\textwidth}}
  \end{minipage} \ec
\vspace{-0.5cm}
  \caption{Another typical light curve}\label{TTT}
\end{figure}
$7 \times 7$ super-pixel shown on figure~(\ref{WWW}). This light curve
first shows the degree of stability ($\simeq$ 0.5\%) we achieve on a
stable pixel. It appears consistent with error bars taken as twice the
photon noise as explained earlier. A second point 
apparent on this light curve is that a magnitude variation $\Delta m
\simeq 0.02$ (800/40000 ADU) of the super-pixel is clearly detectable. 

Figure~\ref{TTT} shows another light curve. On graph ~\ref{TTT}a all frames are
retained, whereas only frames with seeing between $1.2\,''$ and
$1.8\,''$ are kept in graph~\ref{TTT}b.  This illustrates the
instabilities introduced by seeing variations. However, when extreme
seeing have been excluded (graph~\ref{TTT}b), a magnitude change $\Delta m
\simeq 0.015$ is clearly detected. There are two luminosity
variations, therefore 
it is not a microlensing event. 

Plots~\ref{TTT}c and~\ref{TTT}d display the intensity in a square of side 30
elementary pixels (10$\,''$). The hills in the landscape
are structures of M31 and appear in
the same way on both plots. However, a tiny hill at the center of plot
\ref{TTT}c  has grown to a high peak on plot~\ref{TTT}d. The clear
point-spread-function shape of the growing peak tells us that we are
really looking at the variation of the luminosity 
of a star. This is confirmed by the progressive rise and fall of this peak on
 exposures before and after the maximum shown on
 plot~\ref{TTT}d. Clearly,  such a faint variable object would not have been
 detected by monitoring resolved stars.

We have a catalog of a few hundred such variations, most of which
are multiple and therefore are not microlensing events. We are know
working i) to treat the seeing variations ii) to interprete the
variations we see in terms of known types of variable 
stars, iii) to try and isolate events compatible with microlensing and in any
case to evaluate our sensitivity threshold for the detection of
microlensing events. \\

We thank professor A. Gould for useful discussions and suggestions.

\end{document}